\documentclass[aps,preprint,prd,showpacs,nofootinbib]{revtex4-1}
\usepackage{graphicx}
\usepackage{bm}
\usepackage{times}
\usepackage{hyperref}
\usepackage{slashed}
\usepackage{amsmath}
\usepackage{xcolor}
\usepackage{fancyhdr}
\usepackage{footmisc}
\usepackage{soul}
\usepackage{lipsum}
\def\br{\begin{eqnarray}}
\def\er{\end{eqnarray}}
\def\be{\begin{equation}}
\def\ee{\end{equation}}

\def\({\left(}
\def\){\right)}

\def\figureautorefname~#1\null{Fig.\,#1\null}
\def\tableautorefname~#1\null{Tab.\,#1\null}

\def\equationautorefname~#1\null{Eq.\,(#1)\null}

\begin{document}

\title{PandaX-4T limits on $Z^\prime$ mass in 3-3-1LHN model}

\author{Vin\'icius Oliveira$^a$}\email{vlbo@academico.ufpb.br
}  
\author{C. A. de S. Pires$^{a}$}\email{cpires@fisica.ufpb.br} 
\affiliation{$^{a}$Departamento de F\'isica, Universidade Federal da Para\'iba, 
Caixa Postal 5008, 58051-970, Jo\~ao Pessoa - PB, Brazil.}

\date{\today}
\vspace{1cm}
\begin{abstract}
The framework of the  so-called 3-3-1LHN model  accommodates  two different, but viable, scenarios of  dark matter. In one case the dark matter particle is a heavy Dirac neutrino  $N_1$.  In the other case we have a scalar, $\phi$, as a dark matter candidate. In both cases the dark matter phenomenology, relic abundance and scattering cross section off of nuclei are controlled by exchange of $Z^\prime$. We  then investigate the impact on the parameter space $(M_{Z^\prime}\,,\,M_{(N_1\,, \ \phi)})$ due to the recent PandaX-4T experimental result in both scenarios. We obtain that the PandaX-4T experiment  excludes scenarios  with dark matter mass below $1.9$ TeV.  Concerning $Z^\prime$, we find the lower bound  $M_{Z^\prime}> 4.1$ TeV for the case where $N_1$ is the dark matter and $ M_{Z^\prime}>5.7$ TeV for the other case. This implies that the 3-3-1 symmetry is spontaneously broken above $10$ TeV scale. We also comment on the contributions to the relic abundance of  processes involving flavor changing neutral current mediated by $Z^{\prime}$.
\end{abstract}
    
\maketitle
\section{Introduction}

The nature of Dark Matter (DM) continues to be a   challenge to particle physics. At the moment experiments keep making great  effort in trying to decipher its nature
by means of many types of detection \cite{Schnee:2011ooa, Battaglia:2004mp, PhysRevD.79.015008} (direct, indirect and collider) while theorists interpret their results inside  theories that 
pose DM particles. Recently the PandaX-4T \cite{PandaX-4T:2021bab} experiment released its first report concerning dark matter search by means of direct detection. Its null result translates in a stringent limit to the dark matter-nucleon spin independent interactions. 

The 3-3-1 with left-handed neutrino (3-3-1LHN) model is an interesting dark matter model \cite{Dutra:2021lto, Long:2020uyf, Mizukoshi:2010ky, Profumo:2013sca, Cogollo:2014jia, Dong:2015rka, Arcadi:2017xbo, Dong:2013wca, Huong:2011xd, CarcamoHernandez:2019lhv,Alvarez-Salazar:2019npi,Montero:2017yvy,Ferreira:2016uao}. It is based on the  $SU(3)_C \times SU(3)_L \times U(1)_N$ (3-3-1) gauge symmetry \cite{Queiroz:2010rj,Pisano:1992bxx} which poses heavy Dirac neutrinos in its particle spectrum. It carries three DM candidates in its particle spectrum, namely $U^0$, $N_1$ and $\phi$. The first is extremely underabundant while the other two, $N_1$ and $\phi$, are viable DM candidates. Of course they cannot co-exist. A kind of R-parity guarantees the stability of such DM particles. In this work we calculate, for both dark matter cases, the relic abundance including flavor changing neutral process with $Z^{\prime}$ and 
extract bounds on the mass of $Z^{\prime}$ by confronting the theoretical scattering cross section off of  nuclei with the PandaX-4T result.
 
 We organize this work in the following way: in \autoref{sec1} we present the essence of the model; in \autoref{sec3} we investigate the dark matter relic abundance and direct detection experiment; lastly we summarize and draw our conclusions in \autoref{sec4}.

\section{The essence of the 3-3-1LHN Model}
\label{sec1}

In the 3-3-1LHN model the left-handed leptons come in triplet representation, $ f_{a_L}=\left(  \nu_a\,,\, e_{a}\,,\,N_a \right)_L^T$
( $a=1,2,3$ ). For the quark sector, the third generation comes in the triplet representation $Q_{3_L}=\left( u_3\,,\, d_3\,,\, u^{\prime}_{3} \right)_L^T$ while the other two come in an anti-triplet representation of $SU(3)_L$,  $ Q_{i_L}=\left ( d_i\,,\, -u_i \,,\, d^{\prime}_{i} \right)_L$ (i=1,2), as  required by anomaly cancellation. With exception of $\nu_L$, all the left-handed fermions have their right-handed counterpart \cite{Pisano:1992bxx}. The new leptons and quarks are heavy particles with their masses belonging to the 3-3-1 energy scale.  For all these features and the transformations of these fields by the 3-3-1 symmetry, see  \cite{Queiroz:2010rj,Pisano:1992bxx}.

The gauge sector of the model is composed of nine gauge bosons which involves the standard ones  $W^{\pm}$, $Z$ , $\gamma$ and five others called $W^{\prime \pm}$,   $U^0$, $U^{0 \dagger}$, and $Z^{\prime}$. For their masses and  features,  see \cite{Pisano:1992bxx, Long:1995ctv}. The interactions of these gauge bosons with all the fermions of the model are found in TABLE I and Eq.(7) of the paper \cite{Profumo:2013sca}. We use those interactions here.

With the triplet of scalars $\eta=\left( \eta^0\,,\, \eta^-\,,\, \eta^{\prime 0}   \right)^T $, $\rho =\left(\rho^+\,,\, \rho^0\,,\, \rho^{\prime +}  \right)^T$ and $\left(\chi^0\,,\, \chi^-\,,\, \chi^{\prime 0} \right)^T$  the symmetries of the model is correctly broken and mass are  generated  for all particles. The  potential composed with these triplets of scalars   was  developed in \cite{deSPires:2007wat}.

We remark that  the 3-3-1LHN model  has three  viable DM candidates, namely $U^0$, $N$, and $\phi$. Their stability  is guaranteed by a kind of  discrete R-odd parity symmetry $P=(-1)^{3(B-L)+2s}$ where $B$ is baryon number, $L$ is lepton number and $s$ is spin of the corresponding field. By means of this symmetry the model poses  a set of 3-3-1 particles that transform as follows,
\begin{equation} \small
    (N_L\,,\,N_R\,,\, d_i^{\prime}\,,\,u_3^{\prime}\,\,\rho^{\prime +}\,,\,\eta^{\prime 0}\,,\, \chi^0\,,\, \chi^-\,,\, W^{\prime +}\,,\,U^{0\dagger}) \rightarrow -1.
\end{equation}
We refer to these particles as $R$-odd particles. All the other particles of the model transform trivially under $R$-odd parity. Then the electrically neutral  $R$-odd particles $N_1$, $\eta^{\prime 0}$, and $U^0$ are potential dark matter candidates\footnote{We are assuming normal hierarchy among the heavy neutral fermions $N_1\,,\, N_2 \,,\,N_3$ in such a way that $N_1$ is the lightest of them.}. The gauge boson  $U^0$ is  extremely underabundant and naturally discarded as dark matter. Regarding to $N_1$ and $\eta^{\prime 0}$,  they interact one with another by means of the term $\frac{g^{\prime}_{11}v}{2v_{\chi^{\prime}}} \bar \nu_e
N_1 \eta^{\prime 0}$\cite{Mizukoshi:2010ky}.  By enforcing that one of them is the lightest 
$R$-odd particle, then, it gets stable and becomes a good dark matter candidate. In what follows, we obtain the abundance of each candidate, their spin independent cross section and confront them with the recent PandaX-4T.\footnote{From now on we employ the notation of  Ref. \cite{Profumo:2013sca} for all the  fields.}

\section{Relic Abundance and Direct Detection} 
\label{sec3}

The marvelous characteristic of WIMPs is that their interactions manifest at the electroweak scale which naturally leads to the appropriate relic density. Due to this characteristic the WIMP tends to thermalize with the standard model particles in the primordial universe. This happens when its interaction rate is greater than the expansion rate of the Universe. The WIMP decouples when the rate of interactions drops below the expansion rate of the Universe , being cosmologically stable, its abundance keeps constant in the Universe up to today. Additionally, the electroweak scale of WIMPs interaction implies that it is experimentally accessible. Nowadays there are three potential ways to search these particles experimentally \cite{Battaglia:2004mp,Jungman:1995df,Schnee:2011ooa,Cooley:2021rws}, that is: indirectly, directly or collider. In this work we will explore the bounds of the direct detection experiment PandaX-4T \cite{PandaX-4T:2021bab} and, more remarkably, constrain the mass of $Z^\prime$.

\subsection{Relic Abundance}

To obtain the WIMP abundance we need to solve the Boltzmann equation which gives the evolution of the abundance of a generic species in the Universe as a function of the temperature,
\begin{equation}\label{Boltzmann_EQ}
\frac{dY}{dT}=\sqrt{\frac{\pi g_{\ast}(T)}{45}}M_{p} -\left\langle \sigma v\right\rangle(Y^{2}-Y^{2}_{eq})\,,
\end{equation}
where $g_{\ast}$ is the effective number of degrees of freedom, $M_{p}$ is the Planck mass, $Y \equiv n/s$ is the  abundance or number density ($n$) over entropy density ($s$) (while $Y_{eq}$ is the abundance at the equilibrium) and $\left\langle \sigma v\right\rangle$ is the thermally averaged cross section for WIMP annihilation times the relative velocity. 

The particle physics information of the model enters in the  thermally averaged cross section which includes all annihilation and co-annihilation channels. In this work we will assume that $M_{N_1} \ll M_{N_2} \ll M_{N_3} $, which makes the co-annihilation processes irrelevant\footnote{However, even considering degenerate case (as done in \cite{Profumo:2013sca}) the co-annihilation channels are irrelevant.} \cite{PhysRevD.43.3191}. The thermally averaged cross section for annihilation processes ($A+A \to B+B$) is
\begin{equation}       
\langle \sigma v \rangle \equiv \frac{1}{(n_{A}^{eq}(T))^2} \frac{\cal{S}}{32 \left(2 \pi \right)^6} T \int ds \frac{\sqrt{\lambda \left(s, m_{A}^2,m_{A}^2\right)}}{s} \frac{\sqrt{\lambda \left(s, m_{B}^2,m_{B}^2\right)}}{\sqrt{s}} K_1 \left(\frac{\sqrt{s}}{T} \right) \int d \Omega |\mathcal{M}|^2 \,,
\end{equation}
where $\cal{S}$ is the symmetrization factor, $T$ is the thermal bath temperature, $s$ is the Mandelstam variable, $\lambda \left( x, y, z \right)$ is the K{\"a}ll\'en function, $K_1$ is the modified Bessel function of the second kind of order $1$, $\Omega$ is the solid angle between initial and final states in the center of mass frame, and $|\mathcal{M}|^2$ the (not averaged) squared amplitude of the process.

The final relic abundance of a DM candidate is defined to be
\begin{equation}
    \frac{\Omega^0 h^2}{0.11} \simeq \frac{M_{DM}}{1 \ \text{GeV}} \frac{Y^0}{4.34 \times 10^{-10}}\,,
\end{equation}
where $M_{DM}$ represents the DM mass and the label "$^0$" indicates quantities as measured today, with $\Omega^0_{DM} h^2 \simeq 0.11$ being inferred by the Planck satellite \cite{Aghanim:2018eyx}. The $Y^0$ can be obtained by integrating \autoref{Boltzmann_EQ} from $T=T_{0}$ to $T=\infty$, where $T_{0}$ is the temperature of the Universe today.

\begin{figure}[!htb]
\centering
\includegraphics[width=1\columnwidth]{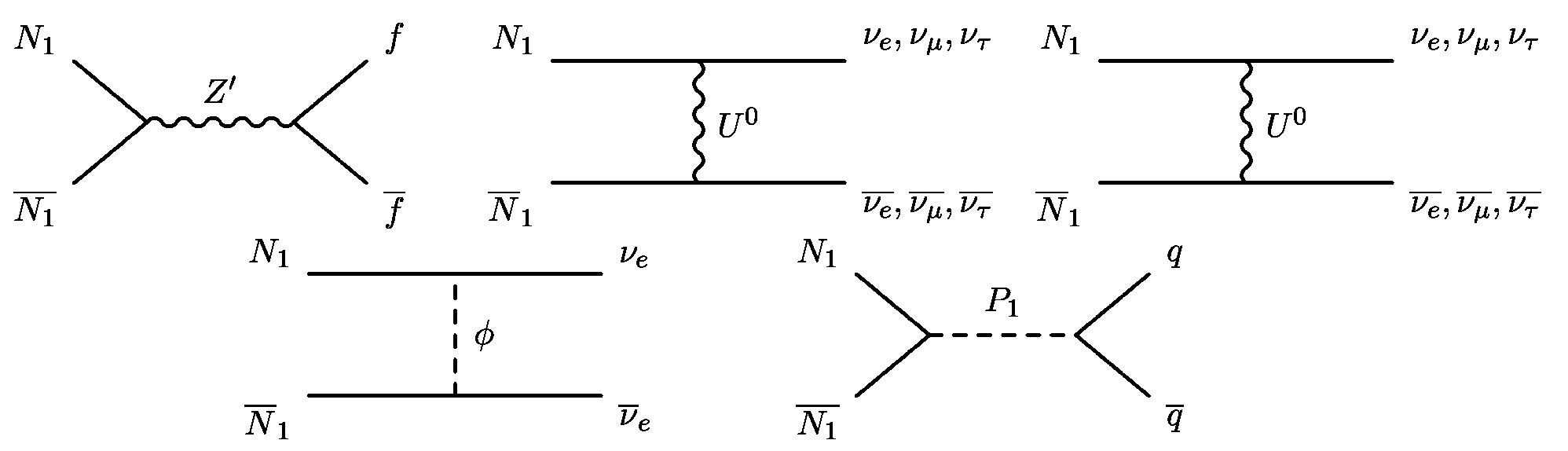}
\caption{The relevant processes that contribute to the abundance of $N_{1}$ with $f$ representing the standard fermions.}
\label{processoN1}
\end{figure}

\begin{figure}[!htb]
\centering
\includegraphics[width=1\columnwidth]{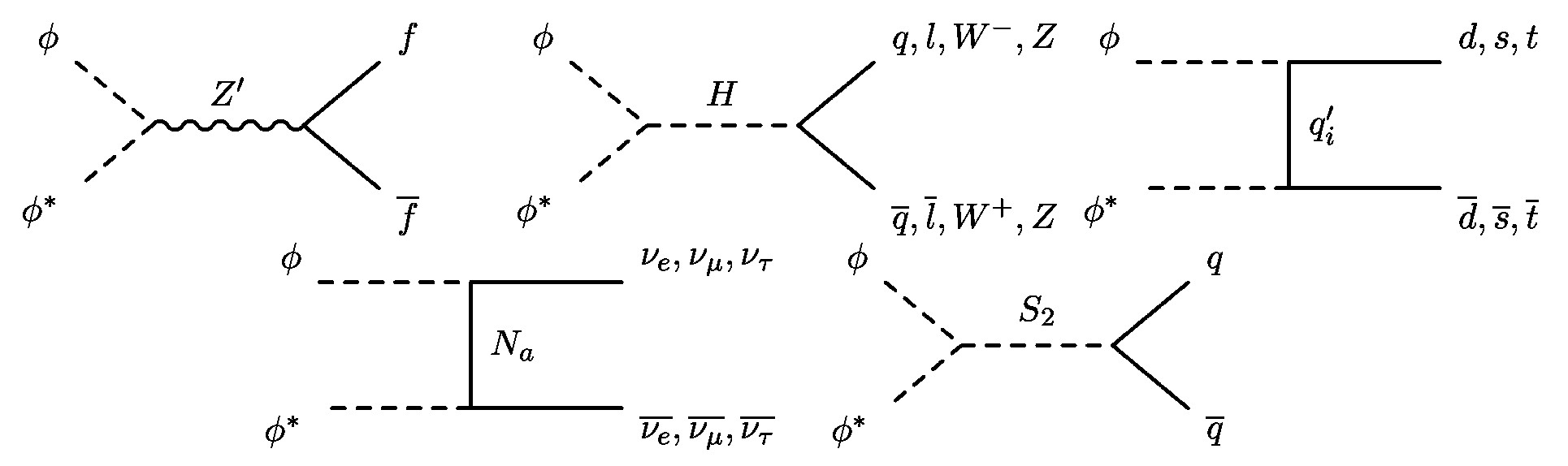}
\caption{The relevant processes that contribute to the abundance of $\phi$ with $f$, $q$ and $l$ representing the standard  quarks and charged leptons, respectively.}
\label{processofi}
\end{figure}

Our results are obtained by using the package micrOMEGAs \cite{Belanger:2018ccd}, which computes the relic density numerically for a given model. The relevant processes which contribute to the abundance of our DM candidates, $N_{1}$ and $\phi$, separately,  are shown in \autoref{processoN1} and \autoref{processofi}. However, other interactions participate in the annihilation process at freeze-out as, for example, flavor changing neutral interactions that we also take into account (where the interactions were obtained from \cite{Long:1999ij}). Then, essentially, we have implemented those interactions in the package CalcHEP \cite{Belyaev:2012qa} that furnishes the model files to be used in micrOMEGAs.

\begin{figure}[!htb]
\centering
\includegraphics[width=0.9\columnwidth]{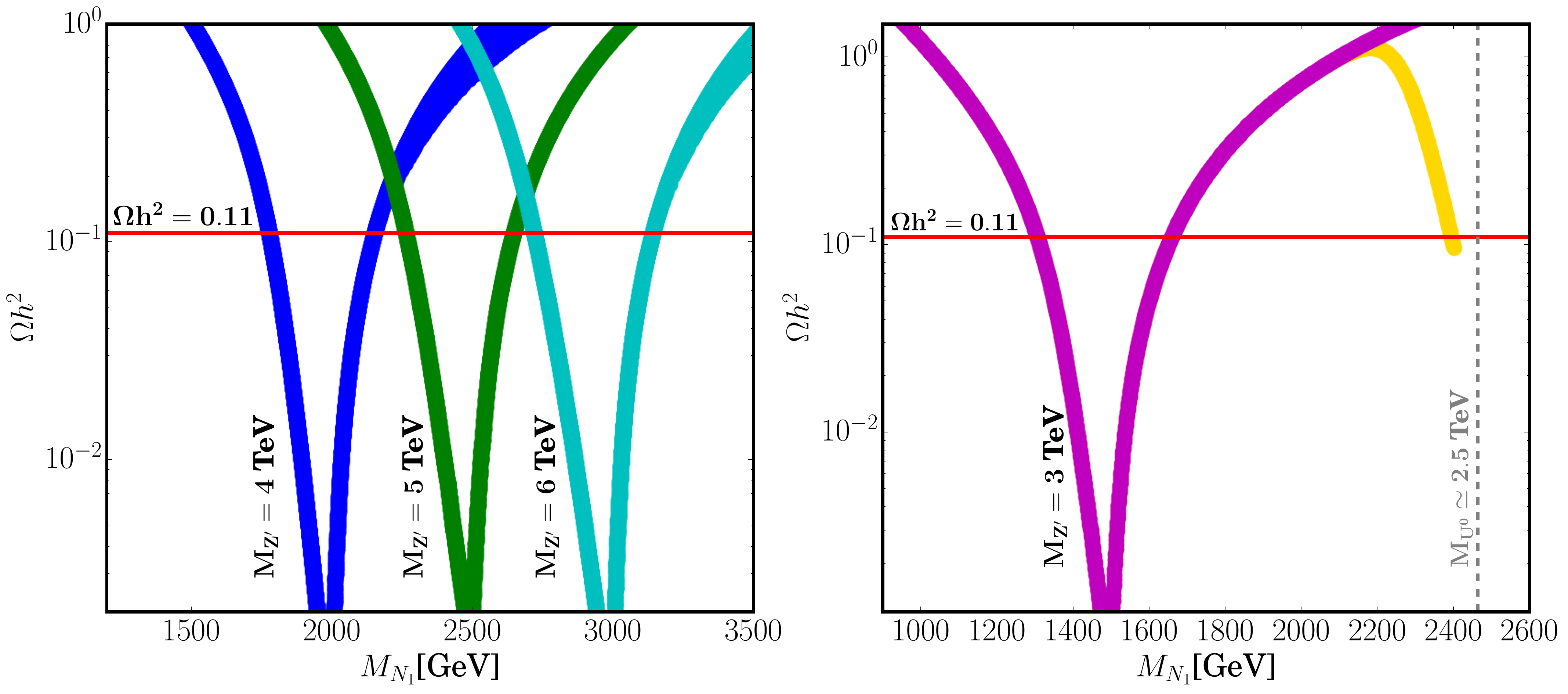}
\caption[]{Left Panel: Relic abundance for the heavy 
neutrino $N_{1}$ for $M_{Z^\prime} = 4$ TeV (blue curve), $5$ TeV (green curve) and  $6$ TeV (cyan curve). Right Panel: Relic abundance for the heavy neutrino $N_{1}$ for $3$ TeV for $\Gamma_{P_1} = \mathcal{O}(1)$ GeV (pink curve) and $\Gamma_{P_1} = \mathcal{O}(10^{-3})$ GeV (yellow curve), showing the $P_1$ resonance. We vary the parameters $\lambda_2$, $\lambda_3$ and $\lambda_6$ given by Eq.(4) of the paper \cite{Profumo:2013sca} always respecting the fact that  $M_H = 125$ GeV (The expression for the Higgs mass involves  $\lambda_2$, $\lambda_3$ and $\lambda_6$ as is shown in Eq.(12) of the paper \cite{Profumo:2013sca}). }
\label{abundanciaN1}
\end{figure}

 Firstly, we will handle the case where $N_1$ is the DM candidate. In the left panel of \autoref{abundanciaN1} we show its relic abundance for $M_{Z^\prime} =$  $4$ TeV (blue curve), $5$ TeV (green curve) and $6$ TeV (cyan curve). We reinforce that in our calculation we took  $ M_{R-odd particles}\gg M_{N_1}$, where $M_{R-odd particles}$ represents the masses of all other $R$-odd particles. In that figure the region in accordance with  the Planck satellite \cite{Aghanim:2018eyx}, $\Omega h^2 = 0.11$, is shown by the red horizontal line. We can observe that the abundance of $N_1$ is suppressed when $M_{N_1} = M_{Z^\prime}/2$, which means the resonance of $Z^\prime$. This fact tells us that the processes mediated by the $Z^\prime$ boson are the most relevant ones. As the reader can see below, direct detection requires $M_{N_1}> 1900$ GeV.   In summary,  for this particular case  $N_1$ fulfills all the conditions to be a dark matter candidate since it is stable and provides the correct abundance of dark matter in the Universe. 
 
The interaction $\bar{\nu}_e N_1 U^0$ requires $M_{N_1} < M_{U^0}$ to guarantee the $N_1$ stability. This implies that always $M_{N_1} < M_{P_1}$. However, for some range of values of $\nu_\chi$ we can have $M_{U^0} \simeq M_{P_1}$ which implies that we can observe the appearance of $P_1$ resonance when $M_{N_1} \simeq M_{P_1}/2$, as is shown in the right panel of  \autoref{abundanciaN1}. The resonance only appears when $\Gamma_{P_1} \sim \mathcal{O}(10^{-3})$ GeV, which is represented by the yellow curve in the right panel of \autoref{abundanciaN1}. However, our case gives $\Gamma_{P_1} \sim \mathcal{O}(1)$ GeV. Consequently, no one resonance can be observed, which is represented by the pink curve of the right panel in \autoref{abundanciaN1}.

\begin{figure}[!htb]
\centering
\includegraphics[width=\columnwidth]{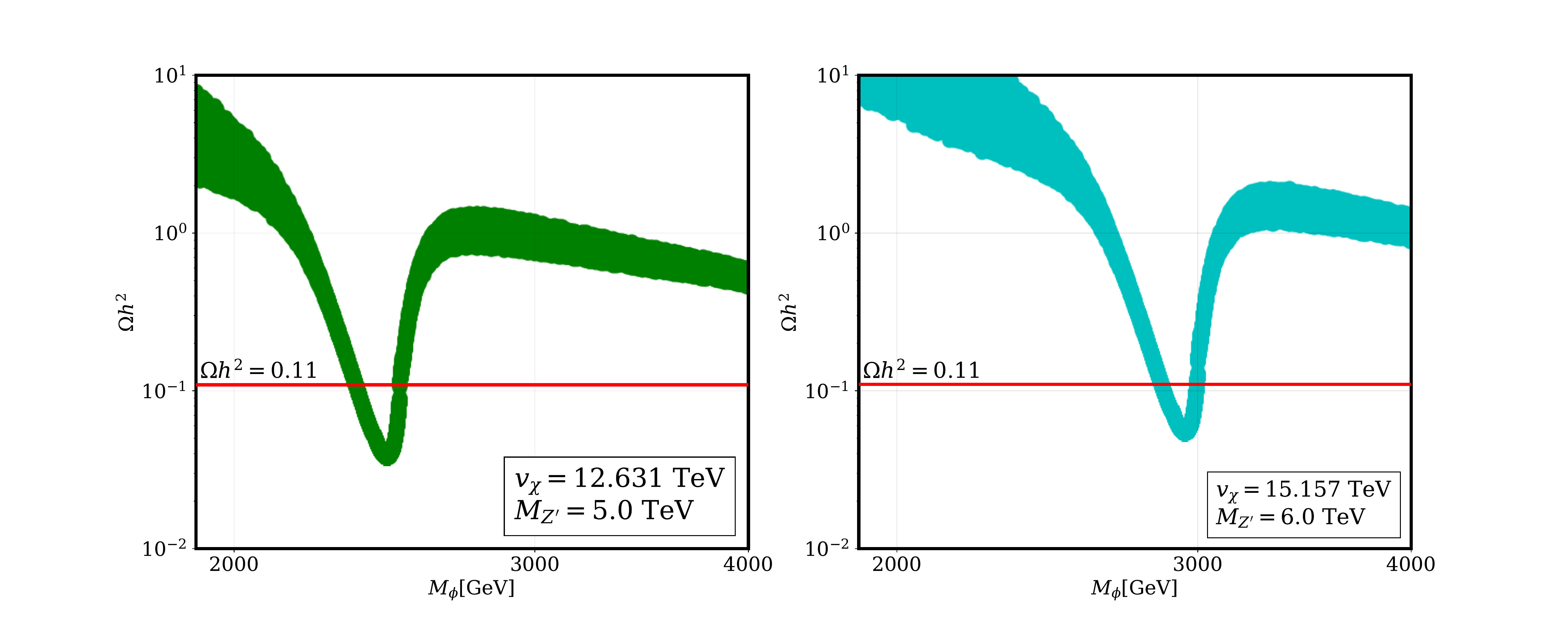}
\caption{The abundance of the scalar $\phi$ for two distinct values of $M_{Z^\prime}$ ($\nu_\chi$). The left panel accounts for $M_{Z^\prime} = 5.0 $ TeV ($\nu_\chi = 12.631$ TeV), and the right panel, for $M_{Z^\prime} = 6.0 $ TeV ($\nu_\chi = 15.157$ TeV).  The variation of $M_\phi$ (given by Eq.(14) of the paper \cite{Profumo:2013sca}) is achieved by varying $\lambda_7$. We also vary the parameters $\lambda_2$, $\lambda_3$ and $\lambda_6$ given by Eq.(4) of the paper \cite{Profumo:2013sca} always respecting the bound imposed by to $M_H = 125$ GeV.}
\label{abundanciaphi}
\end{figure}
Let us now consider the case in which $\phi$ is the lightest $R$-odd particle. In \autoref{abundanciaphi} we show its relic abundance for the cases  $M_{Z^\prime} = 5$ TeV (green curve in left panel) and $6$ TeV (cyan curve in right panel).  The region in accordance with the Planck satellite \cite{Aghanim:2018eyx}, $\Omega h^2 = 0.11$, is shown by the red horizontal line. Observe that the abundance of $\phi$ is suppressed when  $M_\phi = M_{Z^\prime}/2$, which represents the resonance of $Z^\prime$. This reveals to us that the processes mediated by  $Z^\prime$ are the most relevant in this scenario. The reasons for we display $M_\phi$ in \autoref{abundanciaphi} within the range $1.9\,\mbox{TeV} \leq M_\phi \leq 4\,\mbox{TeV}$ are twofold: direct detection exclude light $\phi$ while the trilinear interaction  $\phi H U^0$ imposes that $M_\phi < 4$ TeV (in the left panel) and $M_\phi < 5$ TeV (in the right panel) in order to guarantee the stability of $\phi$. In summary, $\phi$ fulfill all the requisites to constitute the dark matter of the Universe. It is important to note that the coupling of $N_1$  with $Z^\prime$ is approximately twice as large as the coupling of $\phi$ with $Z^\prime$. Because of this the suppression of the abundance of $N_1$ due to $Z^\prime$ resonance tends to be greater than $\phi$.

Thus, we conclude that the model has two viable dark matter candidates and that are exclusive unless degenerate in masses. 
We remark here that the model poses interactions among standard quarks ($q$) and $Z^{\prime}$ that change flavor \cite{Long:1999ij, Liu:1994rx, Rodriguez:2004mw, Benavides:2009cn, Cabarcas:2009vb, Cabarcas:2011hb, Cogollo:2012ek,Machado:2013jca}.  We considered such  contributions in the calculation of the abundance.  As they are suppressed by the quark mixing matrix elements, then their contributions are irrelevant for the abundance in both cases of $N_1$ or $\phi$ as dark matter. \footnote{In regard to lepton flavor violation (LFV) processes like $\mu\rightarrow e \gamma$, once we are considering $N_{1,2,3}$ in a diagonal basis, neither $N_i$ nor $W^{\prime}$ contribute to such  processes. Then we do not need to worry about this here.  }

\subsection{Direct Detection}

The Holy grail of direct dark matter detection is the assumption that the halo of Milk Way is composed by WIMPs, then an infinity of them pass through the Earth's surface each second. 
WIMPs have cross section of approximately weak strength, so it is to be expected that they interact weakly with the Standard Model particles, therefore, as the WIMPs is supposed to pass through the Earth they can be directly detected by their interactions with the material (the nucleons, more precisely the quarks) that compose the detector. The rate of event per unit of time per unit of mass of detector material, which can be simply expressed as \cite{Jungman:1995df}
\begin{equation}
R \simeq \frac{n \langle v \rangle \sigma}{m_N},
\end{equation}
where $\langle v \rangle $ is the average velocity of the incident WIMPs relative to the Earth frame, $\sigma$ is the WIMP-nucleus cross section and $m_N$ is the mass of target nucleus.

The WIMP interacts with the nucleus of the material that composes the detector and deposits an energy $Q$ that is measured \cite{Schnee:2011ooa,Taoso_2008,Hooper:2009zm,Munoz:2003gx,BERTONE2005279,GASCON200496, RAMACHERS2003341,Shan:2007mq}. The WIMPs move in the halo with velocities determined by their velocity distribution function $f(v)$, then the differential scattering event rate can be written
\begin{equation}
    dR = \left( \frac{\rho_0 \sigma_0}{2 m_{DM} \mu_N^2} \right) F^2(Q) \int \frac{f(v)}{v} dv dQ\,,
\end{equation}
where $\rho_0$ is the WIMP density near the Earth, $m_{DM}$ is the DM mass, $\sigma_0$ is the  WIMP-nucleus cross section ignoring the form factor suppression $F(Q)$, $\mu_N = m_{DM} m_N/(m_N + m_{DM})$ is the reduced WIMP-nucleus mass. 

As discussed above, the interactions among dark matter and $Z^{\prime}$ are the most relevant ones. Due to the features of these interactions we will have  two types of WIMP-nucleus interactions: Spin Independent (SI) and Spin Dependent (SD). It is very well known that the SI are the ones we must take into account. Then we will probe here the limits on the SI cross-section of $N_1$ and $\phi$.

The spin independent WIMP-nucleus cross section at zero momentum transfer can be expressed as \cite{Jungman:1995df,Schnee:2011ooa,Belanger:2018ccd}
\be
\sigma_0 = \frac{4 \mu_N^2}{\pi}\left( Z f_p + (A-Z) f_n \right)^2\,,
\label{wimpnucleon}
\ee
where $Z$ is the atomic number, $A$ is the atomic mass and $f_{p}$ and $f_{n}$ are effective couplings with protons and neutrons, respectively, and depend of the particle physics input of a given model. In most cases, the couplings to protons and neutrons are approximately equal $f_{p}\cong f_{n}$, then
\be
\sigma_0 = \sigma^{SI} \frac{\mu_{N}^2}{\mu_n^2 } A^2\,.
\label{wimpnucleoncs}
\ee
where $\mu_{n}$ is the WIMP-nucleon reduced mass and
\begin{equation}\label{eq:sigmaSI}
    \sigma^{SI} = \frac{4 \mu_n^2}{\pi} (f^n)^2\,,
\end{equation}
with $f^n = f_{n,p}$ which is also called WIMP-nucleon amplitude. Experiments tend to constraint the $\sigma^{SI}$, which is nucleus independent.

\begin{figure}[!t]
\centering
\includegraphics[width=0.7\columnwidth]{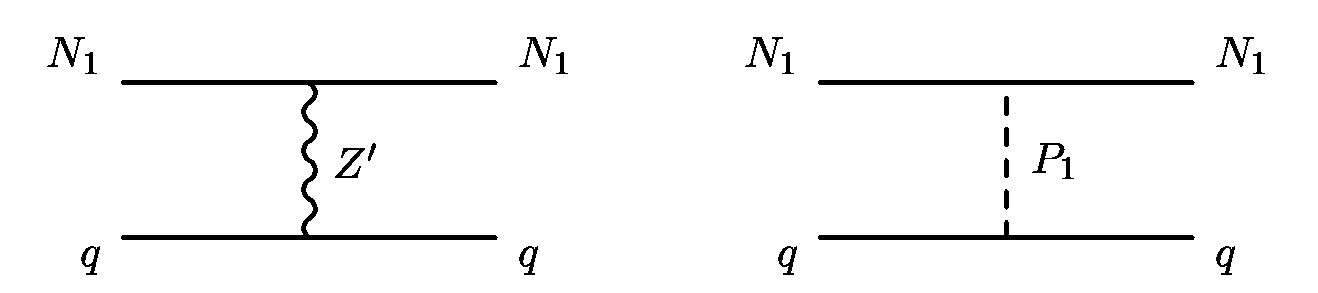}
\caption{Processes which contribute to the WIMP-nucleon cross section of $N_{1}$}
    \label{figN1}
\end{figure}

\begin{figure}[!t]
\centering
\includegraphics[width=1\columnwidth]{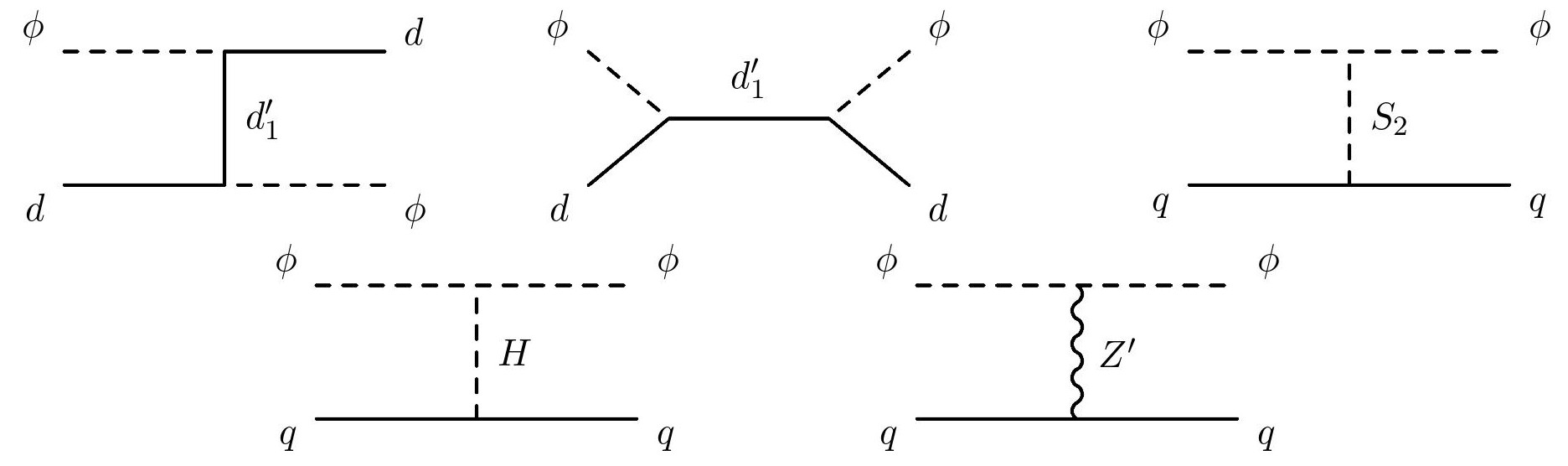}
\caption{Processes which contribute to the WIMP-nucleon cross section of $\phi$.}
\label{figFi}
\end{figure}

The processes which contribute to the spin independent cross section of $N_{1}$ is shown in \autoref{figN1}. In our calculation we took into account both  contributions, however we remark that the process mediated by the pseudo-scalar $P_1$ is completely negligible in comparison to the one mediated by $Z^{\prime}$. This is so because its coupling involves a $\gamma_5$ and is suppressed by a tiny Yukawa coupling, see \cite{Profumo:2013sca}. Additionally, in \autoref{figFi} we have the processes which contribute to spin independent cross section of $\phi$. Our results were obtained by implementing all these interactions in CalcHEP. We made use of micrOMEGAs  to compute $\sigma^{SI}$ through \autoref{eq:sigmaSI}.

After discussing a little bit about the direct detection method and showing the processes which contribute to the WIMP-nucleus cross section, we are able to show and analyze the results for each candidate.

\begin{figure}[!t]
\centering
\includegraphics[width=1\columnwidth]{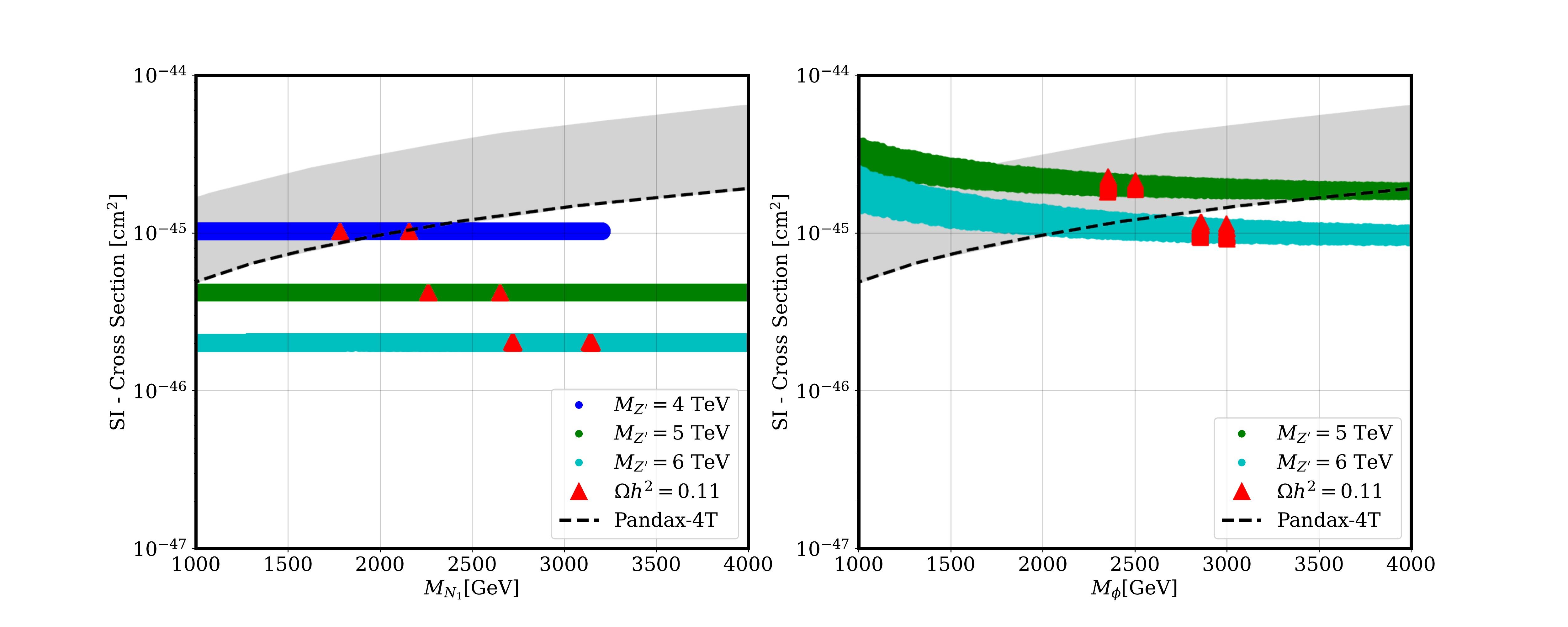}
\caption{The WIMP-nucleon cross section for $N_{1}$ (left panel) and $\phi$ (right panel). The red triangles represents the right DM abundance, for both case. The black dashed line represents the upper limit imposed by the direct detection experiment PandaX-4T  \cite{PandaX-4T:2021bab}. The light grey band represents the $\pm 1 \sigma$ sensitivity band, which is in agreement with Fig.4 of \cite{PandaX-4T:2021bab}.}
\label{DD}
\end{figure}

The main results of this work are shown in \autoref{DD} which consider the recent PandaX-4T result \cite{PandaX-4T:2021bab} (in agreement with Fig.4 of \cite{PandaX-4T:2021bab}). In the left panel of \autoref{DD} we present our numerical results for $N_1$-nucleon cross section as a function of $M_{N_1}$ for $M_{Z^\prime} = 4$ TeV (blue curve), $5$ TeV (green curve) and $6$ TeV (cyan curve). In order to guarantee the stability of $N_1$, due to the trilinear interaction $\nu_e N_1 U^0$, we needed to assume $M_{N_1} < M_{U^0}$. This is the reason why the blue line in the left panel of \autoref{DD} goes up to $M_{N_1} = 3200$ GeV. The black dashed line represents the upper limit imposed by the recent direct detection PandaX-4T experimental result. The region above the dashed line is excluded by the PandaX-4T bound. The red triangles represent the right amount of relic abundance. Then as the red triangles that overlap the blue line lie above the black dashed line, we conclude that the correct abundance and PandaX-4T bound exclude $Z^{\prime}$ with mass of $4$ TeV.

In the right panel of \autoref{DD} we present our numerical results for $\phi$-nucleon cross section as a function of $M_{\phi}$ for two values of $M_{Z^\prime}$. As we can see in this figure, the correct abundance and PandaX-4T bound exclude $Z^{\prime}$ with mass of $5$ TeV for any value of  $ M_\phi$. 

\begin{figure}[!t]
\centering
\includegraphics[width=1\columnwidth]{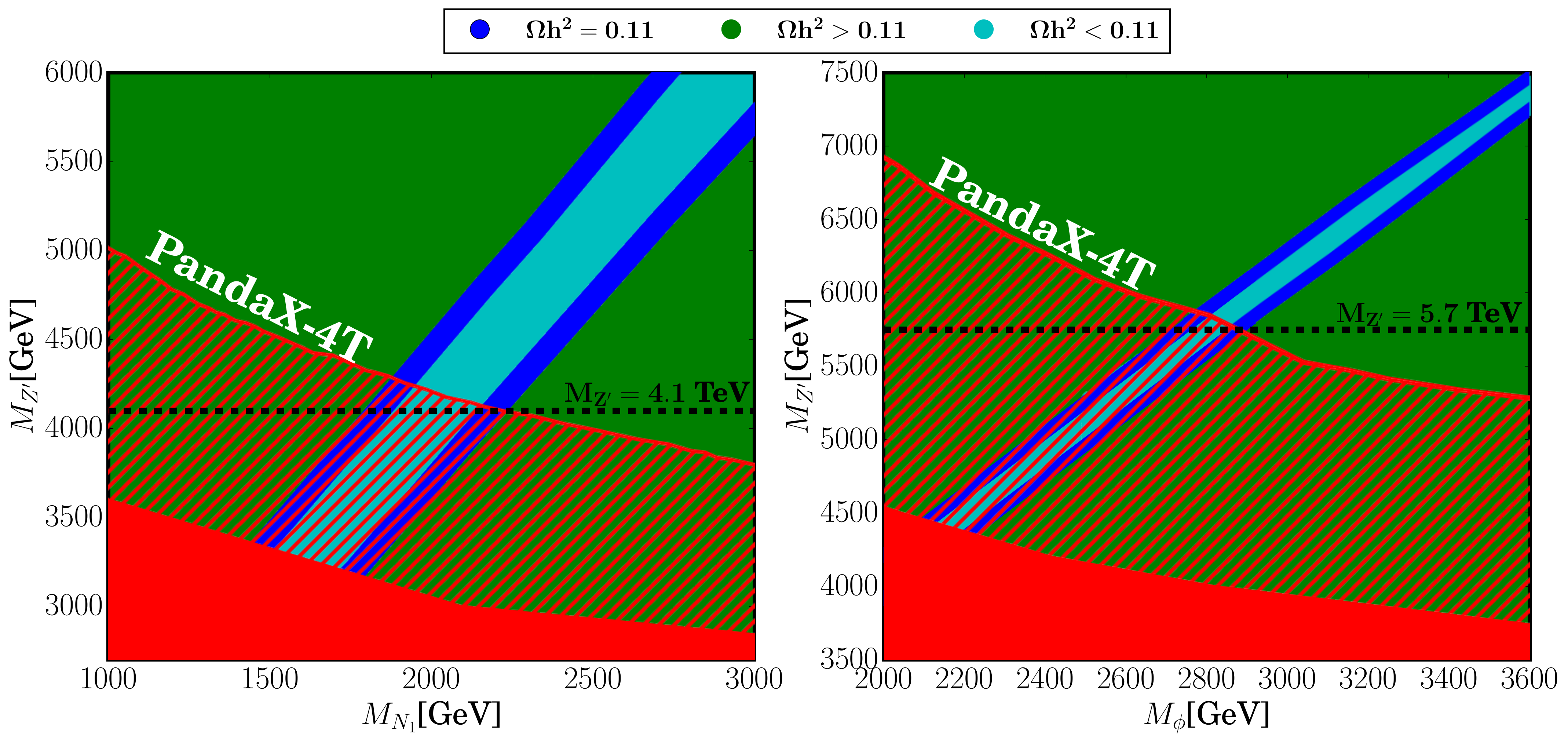}
\caption{Viable parameter space for the DM candidates $N_1$ (left panel) and $\phi$ (right panel) in the plane ($M_{Z^\prime}$, $M_{N_1}$) and ($M_{Z^\prime}$, $M_{\phi}$), respectively. In both panels the correct dark matter relic density ($\Omega h^2 = 0.11$) is represented by blue region, cyan region represents the under-abundance ($\Omega h^2 < 0.11$) parameter space and the green region represents the over-abundance ($\Omega h^2 > 0.11$) parameter space. The red line represents the PandaX-4T direct detection experiment with its $\pm 1 \sigma$ sensitivity band represented by hatched red area. The red region is excluded by the PandaX-4T direct detection experiment.}
\label{ViableParameterspace}
\end{figure}

As noted in both panels of \autoref{DD}, the allowed value of the mass of $Z^{\prime}$ is related to the mass of the dark matter candidate. Then, in \autoref{ViableParameterspace} we present the region of parameter space $(M_{Z^{\prime}}, M_{(N_1, \phi)})$ which is allowed by the recent PandaX-4T. In both figures we can see, represented by the blue region, all the values of the parameter space $(M_{Z^{\prime}}, M_{(N_1, \phi)})$ that respect PandaX-4T and have the correct abundance. Thus we have a lower bound on the mass of $Z^{\prime}$ depending on the mass of the dark matter candidate. For the case in which $N_1$ is the dark matter candidate we have that the lower bound on the mass of $Z^{\prime}$ is $M_{Z^{\prime}}= 4.1$ TeV for $M_{N_1}= 2.2$ GeV. It is represented by the dashed black line in the left panel of \autoref{ViableParameterspace}. For the case where $\phi$ is the dark matter candidate, we have that the lower bound on the mass of $Z^{\prime}$ is $M_{Z^{\prime}}=5.7$ TeV for $M_{\phi}=2.9$ TeV.

\section{Discussion and Conclusions}
\label{sec4}
In this work we investigated the implications of PandaX-4T bound on the parameters of the  dark matter candidates of the 3-3-1LHN model. First of all, PandaX-4T bound is not compatible with light dark matter. In our model, when we assume  that $N_1$ is the dark matter candidate, the PandaX-4T bound requires $M_{N_1}\geq 1.9$ TeV. For the other case where $\phi$ is the dark matter candidate we get $M_\phi \geq 2.8$ TeV.

Another interesting result is that, due to the fact that all relevant processes are mediated by $Z^{\prime}$, PandaX-4T bound may be translated into a lower bound on $Z^{\prime}$ mass but now related to the mass of the dark matter candidate. As shown in the left panel of \autoref{ViableParameterspace} the lower bound on the mass of $Z^{\prime}$ is $M_{Z^{\prime}}\geq 4.1$ TeV ($v_\chi \geq 10357$ GeV) but this requires $M_{N_1}=2200$ GeV. In the case where $\phi$ is the dark matter candidate, this lower bound is yet even more restrictive rendering $ M_{Z^{\prime}} \geq 5.7$ TeV ($v_\chi \geq 14400$ GeV) for $M_\phi=2.9$ TeV. Observe that the bound on $M_{Z^{\prime}}$ is more severe for the $\phi$ scenario. We highlight that the lower bound on $M_{Z^{\prime}}$ obtained here turns out to be more restrictive than  previous works existing in the literature \cite{Profumo:2013sca}, including that imposed by LHC \cite{RamirezBarreto:2010vji,RamirezBarreto:2007cie,RamirezBarreto:2006tn, Coutinho:2013lta}. \footnote{For other source of constraint on the mass of $Z^{\prime}$, see \cite{Long:1999bny}. } Of course we are aware that such a bound on $M_{Z^{\prime}}$ is an implication of the assumption that $N_1$ or $\phi$ is the candidate for the dark matter of the Universe. A recent work on the subject, see \cite{Alves:2022hcp}, found that future hadron colliders  may severely  restrict $M_{Z^{\prime}}$. If this is realized, then , according to \autoref{ViableParameterspace}, $N_1$ and $\phi$ must be heavier in order to give the correct abundance\footnote{We recall that there is upper limit for the WIMP mass that arises from the question of unitarity, see \cite{Griest:1989wd}.}. We finalize saying that  PandaX-4T bound put the dark matter candidates of the 3-3-1LHN model in a scale of energy that can not be probed by the LHC in the present running and pushed the lower bound on the mass of $Z^{\prime}$ for a scale that surpass the existing bound.

\section*{Acknowledgments}
C.A.S.P  was supported by the CNPq research grants No. 304423/2017-3 and V.O was supported by CAPES.

\bibliography{bibliography}

\begin{thebibliography}{49}%
\makeatletter
\providecommand \@ifxundefined [1]{%
 \@ifx{#1\undefined}
}%
\providecommand \@ifnum [1]{%
 \ifnum #1\expandafter \@firstoftwo
 \else \expandafter \@secondoftwo
 \fi
}%
\providecommand \@ifx [1]{%
 \ifx #1\expandafter \@firstoftwo
 \else \expandafter \@secondoftwo
 \fi
}%
\providecommand \natexlab [1]{#1}%
\providecommand \enquote  [1]{``#1''}%
\providecommand \bibnamefont  [1]{#1}%
\providecommand \bibfnamefont [1]{#1}%
\providecommand \citenamefont [1]{#1}%
\providecommand \href@noop [0]{\@secondoftwo}%
\providecommand \href [0]{\begingroup \@sanitize@url \@href}%
\providecommand \@href[1]{\@@startlink{#1}\@@href}%
\providecommand \@@href[1]{\endgroup#1\@@endlink}%
\providecommand \@sanitize@url [0]{\catcode `\\12\catcode `\$12\catcode
  `\&12\catcode `\#12\catcode `\^12\catcode `\_12\catcode `\%12\relax}%
\providecommand \@@startlink[1]{}%
\providecommand \@@endlink[0]{}%
\providecommand \url  [0]{\begingroup\@sanitize@url \@url }%
\providecommand \@url [1]{\endgroup\@href {#1}{\urlprefix }}%
\providecommand \urlprefix  [0]{URL }%
\providecommand \Eprint [0]{\href }%
\providecommand \doibase [0]{http://dx.doi.org/}%
\providecommand \selectlanguage [0]{\@gobble}%
\providecommand \bibinfo  [0]{\@secondoftwo}%
\providecommand \bibfield  [0]{\@secondoftwo}%
\providecommand \translation [1]{[#1]}%
\providecommand \BibitemOpen [0]{}%
\providecommand \bibitemStop [0]{}%
\providecommand \bibitemNoStop [0]{.\EOS\space}%
\providecommand \EOS [0]{\spacefactor3000\relax}%
\providecommand \BibitemShut  [1]{\csname bibitem#1\endcsname}%
\let\auto@bib@innerbib\@empty
\bibitem [{\citenamefont {Schnee}(2011)}]{Schnee:2011ooa}%
  \BibitemOpen
  \bibfield  {author} {\bibinfo {author} {\bibfnamefont {R.~W.}\ \bibnamefont
  {Schnee}},\ }in\ \href {\doibase 10.1142/9789814327183_0014} {\emph {\bibinfo
  {booktitle} {{Theoretical Advanced Study Institute in Elementary Particle
  Physics}: {Physics of the Large and the Small}}}}\ (\bibinfo {year} {2011})\
  pp.\ \bibinfo {pages} {775--829},\ \Eprint {http://arxiv.org/abs/1101.5205}
  {arXiv:1101.5205 [astro-ph.CO]} \BibitemShut {NoStop}%
\bibitem [{\citenamefont {Battaglia}\ \emph {et~al.}(2004)\citenamefont
  {Battaglia}, \citenamefont {Hinchliffe},\ and\ \citenamefont
  {Tovey}}]{Battaglia:2004mp}%
  \BibitemOpen
  \bibfield  {author} {\bibinfo {author} {\bibfnamefont {M.}~\bibnamefont
  {Battaglia}}, \bibinfo {author} {\bibfnamefont {I.}~\bibnamefont
  {Hinchliffe}}, \ and\ \bibinfo {author} {\bibfnamefont {D.}~\bibnamefont
  {Tovey}},\ }\href {\doibase 10.1088/0954-3899/30/10/R01} {\bibfield
  {journal} {\bibinfo  {journal} {J. Phys. G}\ }\textbf {\bibinfo {volume}
  {30}},\ \bibinfo {pages} {R217} (\bibinfo {year} {2004})},\ \Eprint
  {http://arxiv.org/abs/hep-ph/0406147} {arXiv:hep-ph/0406147} \BibitemShut
  {NoStop}%
\bibitem [{\citenamefont {B\'elanger}\ \emph {et~al.}(2009)\citenamefont
  {B\'elanger}, \citenamefont {Nezri},\ and\ \citenamefont
  {Pukhov}}]{PhysRevD.79.015008}%
  \BibitemOpen
  \bibfield  {author} {\bibinfo {author} {\bibfnamefont {G.}~\bibnamefont
  {B\'elanger}}, \bibinfo {author} {\bibfnamefont {E.}~\bibnamefont {Nezri}}, \
  and\ \bibinfo {author} {\bibfnamefont {A.}~\bibnamefont {Pukhov}},\ }\href
  {\doibase 10.1103/PhysRevD.79.015008} {\bibfield  {journal} {\bibinfo
  {journal} {Phys. Rev. D}\ }\textbf {\bibinfo {volume} {79}},\ \bibinfo
  {pages} {015008} (\bibinfo {year} {2009})}\BibitemShut {NoStop}%
\bibitem [{\citenamefont {Meng}\ \emph {et~al.}(2021)\citenamefont {Meng} \emph
  {et~al.}}]{PandaX-4T:2021bab}%
  \BibitemOpen
  \bibfield  {author} {\bibinfo {author} {\bibfnamefont {Y.}~\bibnamefont
  {Meng}} \emph {et~al.} (\bibinfo {collaboration} {PandaX-4T}),\ }\href
  {\doibase 10.1103/PhysRevLett.127.261802} {\bibfield  {journal} {\bibinfo
  {journal} {Phys. Rev. Lett.}\ }\textbf {\bibinfo {volume} {127}},\ \bibinfo
  {pages} {261802} (\bibinfo {year} {2021})},\ \Eprint
  {http://arxiv.org/abs/2107.13438} {arXiv:2107.13438 [hep-ex]} \BibitemShut
  {NoStop}%
\bibitem [{\citenamefont {Dutra}\ \emph {et~al.}(2021)\citenamefont {Dutra},
  \citenamefont {Oliveira}, \citenamefont {de~S.~Pires},\ and\ \citenamefont
  {Queiroz}}]{Dutra:2021lto}%
  \BibitemOpen
  \bibfield  {author} {\bibinfo {author} {\bibfnamefont {M.}~\bibnamefont
  {Dutra}}, \bibinfo {author} {\bibfnamefont {V.}~\bibnamefont {Oliveira}},
  \bibinfo {author} {\bibfnamefont {C.~A.}\ \bibnamefont {de~S.~Pires}}, \ and\
  \bibinfo {author} {\bibfnamefont {F.~S.}\ \bibnamefont {Queiroz}},\ }\href
  {\doibase 10.1007/JHEP10(2021)005} {\bibfield  {journal} {\bibinfo  {journal}
  {JHEP}\ }\textbf {\bibinfo {volume} {10}},\ \bibinfo {pages} {005} (\bibinfo
  {year} {2021})},\ \Eprint {http://arxiv.org/abs/2104.14542} {arXiv:2104.14542
  [hep-ph]} \BibitemShut {NoStop}%
\bibitem [{\citenamefont {Long}\ \emph {et~al.}(2020)\citenamefont {Long},
  \citenamefont {Soa}, \citenamefont {Binh},\ and\ \citenamefont
  {C\'arcamo~Hern\'andez}}]{Long:2020uyf}%
  \BibitemOpen
  \bibfield  {author} {\bibinfo {author} {\bibfnamefont {H.~N.}\ \bibnamefont
  {Long}}, \bibinfo {author} {\bibfnamefont {D.~V.}\ \bibnamefont {Soa}},
  \bibinfo {author} {\bibfnamefont {V.~H.}\ \bibnamefont {Binh}}, \ and\
  \bibinfo {author} {\bibfnamefont {A.~E.}\ \bibnamefont
  {C\'arcamo~Hern\'andez}},\ }\href@noop {} {\  (\bibinfo {year} {2020})},\
  \Eprint {http://arxiv.org/abs/2007.05004} {arXiv:2007.05004 [hep-ph]}
  \BibitemShut {NoStop}%
\bibitem [{\citenamefont {Mizukoshi}\ \emph {et~al.}(2011)\citenamefont
  {Mizukoshi}, \citenamefont {de~S.~Pires}, \citenamefont {Queiroz},\ and\
  \citenamefont {Rodrigues~da Silva}}]{Mizukoshi:2010ky}%
  \BibitemOpen
  \bibfield  {author} {\bibinfo {author} {\bibfnamefont {J.~K.}\ \bibnamefont
  {Mizukoshi}}, \bibinfo {author} {\bibfnamefont {C.~A.}\ \bibnamefont
  {de~S.~Pires}}, \bibinfo {author} {\bibfnamefont {F.~S.}\ \bibnamefont
  {Queiroz}}, \ and\ \bibinfo {author} {\bibfnamefont {P.~S.}\ \bibnamefont
  {Rodrigues~da Silva}},\ }\href {\doibase 10.1103/PhysRevD.83.065024}
  {\bibfield  {journal} {\bibinfo  {journal} {Phys. Rev. D}\ }\textbf {\bibinfo
  {volume} {83}},\ \bibinfo {pages} {065024} (\bibinfo {year} {2011})},\
  \Eprint {http://arxiv.org/abs/1010.4097} {arXiv:1010.4097 [hep-ph]}
  \BibitemShut {NoStop}%
\bibitem [{\citenamefont {Profumo}\ and\ \citenamefont
  {Queiroz}(2014)}]{Profumo:2013sca}%
  \BibitemOpen
  \bibfield  {author} {\bibinfo {author} {\bibfnamefont {S.}~\bibnamefont
  {Profumo}}\ and\ \bibinfo {author} {\bibfnamefont {F.~S.}\ \bibnamefont
  {Queiroz}},\ }\href {\doibase 10.1140/epjc/s10052-014-2960-x} {\bibfield
  {journal} {\bibinfo  {journal} {Eur. Phys. J. C}\ }\textbf {\bibinfo {volume}
  {74}},\ \bibinfo {pages} {2960} (\bibinfo {year} {2014})},\ \Eprint
  {http://arxiv.org/abs/1307.7802} {arXiv:1307.7802 [hep-ph]} \BibitemShut
  {NoStop}%
\bibitem [{\citenamefont {Cogollo}\ \emph {et~al.}(2014)\citenamefont
  {Cogollo}, \citenamefont {Gonzalez-Morales}, \citenamefont {Queiroz},\ and\
  \citenamefont {Teles}}]{Cogollo:2014jia}%
  \BibitemOpen
  \bibfield  {author} {\bibinfo {author} {\bibfnamefont {D.}~\bibnamefont
  {Cogollo}}, \bibinfo {author} {\bibfnamefont {A.~X.}\ \bibnamefont
  {Gonzalez-Morales}}, \bibinfo {author} {\bibfnamefont {F.~S.}\ \bibnamefont
  {Queiroz}}, \ and\ \bibinfo {author} {\bibfnamefont {P.~R.}\ \bibnamefont
  {Teles}},\ }\href {\doibase 10.1088/1475-7516/2014/11/002} {\bibfield
  {journal} {\bibinfo  {journal} {JCAP}\ }\textbf {\bibinfo {volume} {11}},\
  \bibinfo {pages} {002} (\bibinfo {year} {2014})},\ \Eprint
  {http://arxiv.org/abs/1402.3271} {arXiv:1402.3271 [hep-ph]} \BibitemShut
  {NoStop}%
\bibitem [{\citenamefont {Dong}\ \emph {et~al.}(2015)\citenamefont {Dong},
  \citenamefont {Kim}, \citenamefont {Soa},\ and\ \citenamefont
  {Thuy}}]{Dong:2015rka}%
  \BibitemOpen
  \bibfield  {author} {\bibinfo {author} {\bibfnamefont {P.~V.}\ \bibnamefont
  {Dong}}, \bibinfo {author} {\bibfnamefont {C.~S.}\ \bibnamefont {Kim}},
  \bibinfo {author} {\bibfnamefont {D.~V.}\ \bibnamefont {Soa}}, \ and\
  \bibinfo {author} {\bibfnamefont {N.~T.}\ \bibnamefont {Thuy}},\ }\href
  {\doibase 10.1103/PhysRevD.91.115019} {\bibfield  {journal} {\bibinfo
  {journal} {Phys. Rev. D}\ }\textbf {\bibinfo {volume} {91}},\ \bibinfo
  {pages} {115019} (\bibinfo {year} {2015})},\ \Eprint
  {http://arxiv.org/abs/1501.04385} {arXiv:1501.04385 [hep-ph]} \BibitemShut
  {NoStop}%
\bibitem [{\citenamefont {Arcadi}\ \emph {et~al.}(2018)\citenamefont {Arcadi},
  \citenamefont {Ferreira}, \citenamefont {Goertz}, \citenamefont {Guzzo},
  \citenamefont {Queiroz},\ and\ \citenamefont {Santos}}]{Arcadi:2017xbo}%
  \BibitemOpen
  \bibfield  {author} {\bibinfo {author} {\bibfnamefont {G.}~\bibnamefont
  {Arcadi}}, \bibinfo {author} {\bibfnamefont {C.~P.}\ \bibnamefont
  {Ferreira}}, \bibinfo {author} {\bibfnamefont {F.}~\bibnamefont {Goertz}},
  \bibinfo {author} {\bibfnamefont {M.~M.}\ \bibnamefont {Guzzo}}, \bibinfo
  {author} {\bibfnamefont {F.~S.}\ \bibnamefont {Queiroz}}, \ and\ \bibinfo
  {author} {\bibfnamefont {A.~C.~O.}\ \bibnamefont {Santos}},\ }\href {\doibase
  10.1103/PhysRevD.97.075022} {\bibfield  {journal} {\bibinfo  {journal} {Phys.
  Rev. D}\ }\textbf {\bibinfo {volume} {97}},\ \bibinfo {pages} {075022}
  (\bibinfo {year} {2018})},\ \Eprint {http://arxiv.org/abs/1712.02373}
  {arXiv:1712.02373 [hep-ph]} \BibitemShut {NoStop}%
\bibitem [{\citenamefont {Dong}\ \emph {et~al.}(2013)\citenamefont {Dong},
  \citenamefont {Hung},\ and\ \citenamefont {Tham}}]{Dong:2013wca}%
  \BibitemOpen
  \bibfield  {author} {\bibinfo {author} {\bibfnamefont {P.~V.}\ \bibnamefont
  {Dong}}, \bibinfo {author} {\bibfnamefont {H.~T.}\ \bibnamefont {Hung}}, \
  and\ \bibinfo {author} {\bibfnamefont {T.~D.}\ \bibnamefont {Tham}},\ }\href
  {\doibase 10.1103/PhysRevD.87.115003} {\bibfield  {journal} {\bibinfo
  {journal} {Phys. Rev. D}\ }\textbf {\bibinfo {volume} {87}},\ \bibinfo
  {pages} {115003} (\bibinfo {year} {2013})},\ \Eprint
  {http://arxiv.org/abs/1305.0369} {arXiv:1305.0369 [hep-ph]} \BibitemShut
  {NoStop}%
\bibitem [{\citenamefont {Huong}\ \emph {et~al.}(2011)\citenamefont {Huong},
  \citenamefont {Kim}, \citenamefont {Long},\ and\ \citenamefont
  {Thuy}}]{Huong:2011xd}%
  \BibitemOpen
  \bibfield  {author} {\bibinfo {author} {\bibfnamefont {D.~T.}\ \bibnamefont
  {Huong}}, \bibinfo {author} {\bibfnamefont {C.~S.}\ \bibnamefont {Kim}},
  \bibinfo {author} {\bibfnamefont {H.~N.}\ \bibnamefont {Long}}, \ and\
  \bibinfo {author} {\bibfnamefont {N.~T.}\ \bibnamefont {Thuy}},\ }\href@noop
  {} {\  (\bibinfo {year} {2011})},\ \Eprint {http://arxiv.org/abs/1110.1482}
  {arXiv:1110.1482 [hep-ph]} \BibitemShut {NoStop}%
\bibitem [{\citenamefont {C\'arcamo~Hern\'andez}\ \emph
  {et~al.}(2020)\citenamefont {C\'arcamo~Hern\'andez}, \citenamefont {Huong},\
  and\ \citenamefont {Long}}]{CarcamoHernandez:2019lhv}%
  \BibitemOpen
  \bibfield  {author} {\bibinfo {author} {\bibfnamefont {A.~E.}\ \bibnamefont
  {C\'arcamo~Hern\'andez}}, \bibinfo {author} {\bibfnamefont {D.~T.}\
  \bibnamefont {Huong}}, \ and\ \bibinfo {author} {\bibfnamefont {H.~N.}\
  \bibnamefont {Long}},\ }\href {\doibase 10.1103/PhysRevD.102.055002}
  {\bibfield  {journal} {\bibinfo  {journal} {Phys. Rev. D}\ }\textbf {\bibinfo
  {volume} {102}},\ \bibinfo {pages} {055002} (\bibinfo {year} {2020})},\
  \Eprint {http://arxiv.org/abs/1910.12877} {arXiv:1910.12877 [hep-ph]}
  \BibitemShut {NoStop}%
\bibitem [{\citenamefont {Alvarez-Salazar}\ \emph {et~al.}(2019)\citenamefont
  {Alvarez-Salazar}, \citenamefont {Peres},\ and\ \citenamefont
  {S\'anchez-Vega}}]{Alvarez-Salazar:2019npi}%
  \BibitemOpen
  \bibfield  {author} {\bibinfo {author} {\bibfnamefont {C.~E.}\ \bibnamefont
  {Alvarez-Salazar}}, \bibinfo {author} {\bibfnamefont {O.~L.~G.}\ \bibnamefont
  {Peres}}, \ and\ \bibinfo {author} {\bibfnamefont {B.~L.}\ \bibnamefont
  {S\'anchez-Vega}},\ }\href {\doibase 10.1002/asna.201913577} {\bibfield
  {journal} {\bibinfo  {journal} {Astron. Nachr.}\ }\textbf {\bibinfo {volume}
  {340}},\ \bibinfo {pages} {135} (\bibinfo {year} {2019})}\BibitemShut
  {NoStop}%
\bibitem [{\citenamefont {Montero}\ \emph {et~al.}(2018)\citenamefont
  {Montero}, \citenamefont {Romero},\ and\ \citenamefont
  {S\'anchez-Vega}}]{Montero:2017yvy}%
  \BibitemOpen
  \bibfield  {author} {\bibinfo {author} {\bibfnamefont {J.~C.}\ \bibnamefont
  {Montero}}, \bibinfo {author} {\bibfnamefont {A.}~\bibnamefont {Romero}}, \
  and\ \bibinfo {author} {\bibfnamefont {B.~L.}\ \bibnamefont
  {S\'anchez-Vega}},\ }\href {\doibase 10.1103/PhysRevD.97.063015} {\bibfield
  {journal} {\bibinfo  {journal} {Phys. Rev. D}\ }\textbf {\bibinfo {volume}
  {97}},\ \bibinfo {pages} {063015} (\bibinfo {year} {2018})},\ \Eprint
  {http://arxiv.org/abs/1709.04535} {arXiv:1709.04535 [hep-ph]} \BibitemShut
  {NoStop}%
\bibitem [{\citenamefont {Ferreira}\ \emph {et~al.}(2017)\citenamefont
  {Ferreira}, \citenamefont {de~S.~Pires}, \citenamefont {Rodrigues},\ and\
  \citenamefont {Rodrigues~da Silva}}]{Ferreira:2016uao}%
  \BibitemOpen
  \bibfield  {author} {\bibinfo {author} {\bibfnamefont {J.~G.}\ \bibnamefont
  {Ferreira}}, \bibinfo {author} {\bibfnamefont {C.~A.}\ \bibnamefont
  {de~S.~Pires}}, \bibinfo {author} {\bibfnamefont {J.~G.}\ \bibnamefont
  {Rodrigues}}, \ and\ \bibinfo {author} {\bibfnamefont {P.~S.}\ \bibnamefont
  {Rodrigues~da Silva}},\ }\href {\doibase 10.1016/j.physletb.2017.05.034}
  {\bibfield  {journal} {\bibinfo  {journal} {Phys. Lett. B}\ }\textbf
  {\bibinfo {volume} {771}},\ \bibinfo {pages} {199} (\bibinfo {year}
  {2017})},\ \Eprint {http://arxiv.org/abs/1612.01463} {arXiv:1612.01463
  [hep-ph]} \BibitemShut {NoStop}%
\bibitem [{\citenamefont {Queiroz}\ \emph {et~al.}(2010)\citenamefont
  {Queiroz}, \citenamefont {de~S.~Pires},\ and\ \citenamefont
  {da~Silva}}]{Queiroz:2010rj}%
  \BibitemOpen
  \bibfield  {author} {\bibinfo {author} {\bibfnamefont {F.}~\bibnamefont
  {Queiroz}}, \bibinfo {author} {\bibfnamefont {C.~A.}\ \bibnamefont
  {de~S.~Pires}}, \ and\ \bibinfo {author} {\bibfnamefont {P.~S.~R.}\
  \bibnamefont {da~Silva}},\ }\href {\doibase 10.1103/PhysRevD.82.065018}
  {\bibfield  {journal} {\bibinfo  {journal} {Phys. Rev. D}\ }\textbf {\bibinfo
  {volume} {82}},\ \bibinfo {pages} {065018} (\bibinfo {year} {2010})},\
  \Eprint {http://arxiv.org/abs/1003.1270} {arXiv:1003.1270 [hep-ph]}
  \BibitemShut {NoStop}%
\bibitem [{\citenamefont {Pisano}\ and\ \citenamefont
  {Pleitez}(1992)}]{Pisano:1992bxx}%
  \BibitemOpen
  \bibfield  {author} {\bibinfo {author} {\bibfnamefont {F.}~\bibnamefont
  {Pisano}}\ and\ \bibinfo {author} {\bibfnamefont {V.}~\bibnamefont
  {Pleitez}},\ }\href {\doibase 10.1103/PhysRevD.46.410} {\bibfield  {journal}
  {\bibinfo  {journal} {Phys. Rev. D}\ }\textbf {\bibinfo {volume} {46}},\
  \bibinfo {pages} {410} (\bibinfo {year} {1992})},\ \Eprint
  {http://arxiv.org/abs/hep-ph/9206242} {arXiv:hep-ph/9206242} \BibitemShut
  {NoStop}%
\bibitem [{\citenamefont {Long}(1996)}]{Long:1995ctv}%
  \BibitemOpen
  \bibfield  {author} {\bibinfo {author} {\bibfnamefont {H.~N.}\ \bibnamefont
  {Long}},\ }\href {\doibase 10.1103/PhysRevD.53.437} {\bibfield  {journal}
  {\bibinfo  {journal} {Phys. Rev. D}\ }\textbf {\bibinfo {volume} {53}},\
  \bibinfo {pages} {437} (\bibinfo {year} {1996})},\ \Eprint
  {http://arxiv.org/abs/hep-ph/9504274} {arXiv:hep-ph/9504274} \BibitemShut
  {NoStop}%
\bibitem [{\citenamefont {de~S.~Pires}\ and\ \citenamefont {Rodrigues~da
  Silva}(2007)}]{deSPires:2007wat}%
  \BibitemOpen
  \bibfield  {author} {\bibinfo {author} {\bibfnamefont {C.~A.}\ \bibnamefont
  {de~S.~Pires}}\ and\ \bibinfo {author} {\bibfnamefont {P.~S.}\ \bibnamefont
  {Rodrigues~da Silva}},\ }\href {\doibase 10.1088/1475-7516/2007/12/012}
  {\bibfield  {journal} {\bibinfo  {journal} {JCAP}\ }\textbf {\bibinfo
  {volume} {12}},\ \bibinfo {pages} {012} (\bibinfo {year} {2007})},\ \Eprint
  {http://arxiv.org/abs/0710.2104} {arXiv:0710.2104 [hep-ph]} \BibitemShut
  {NoStop}%
\bibitem [{\citenamefont {Jungman}\ \emph {et~al.}(1996)\citenamefont
  {Jungman}, \citenamefont {Kamionkowski},\ and\ \citenamefont
  {Griest}}]{Jungman:1995df}%
  \BibitemOpen
  \bibfield  {author} {\bibinfo {author} {\bibfnamefont {G.}~\bibnamefont
  {Jungman}}, \bibinfo {author} {\bibfnamefont {M.}~\bibnamefont
  {Kamionkowski}}, \ and\ \bibinfo {author} {\bibfnamefont {K.}~\bibnamefont
  {Griest}},\ }\href {\doibase 10.1016/0370-1573(95)00058-5} {\bibfield
  {journal} {\bibinfo  {journal} {Phys. Rept.}\ }\textbf {\bibinfo {volume}
  {267}},\ \bibinfo {pages} {195} (\bibinfo {year} {1996})},\ \Eprint
  {http://arxiv.org/abs/hep-ph/9506380} {arXiv:hep-ph/9506380} \BibitemShut
  {NoStop}%
\bibitem [{\citenamefont {Cooley}(2021)}]{Cooley:2021rws}%
  \BibitemOpen
  \bibfield  {author} {\bibinfo {author} {\bibfnamefont {J.}~\bibnamefont
  {Cooley}},\ }in\ \href@noop {} {\emph {\bibinfo {booktitle} {{Les Houches
  summer school on Dark Matter}}}}\ (\bibinfo {year} {2021})\ \Eprint
  {http://arxiv.org/abs/2110.02359} {arXiv:2110.02359 [hep-ph]} \BibitemShut
  {NoStop}%
\bibitem [{\citenamefont {Griest}\ and\ \citenamefont
  {Seckel}(1991)}]{PhysRevD.43.3191}%
  \BibitemOpen
  \bibfield  {author} {\bibinfo {author} {\bibfnamefont {K.}~\bibnamefont
  {Griest}}\ and\ \bibinfo {author} {\bibfnamefont {D.}~\bibnamefont
  {Seckel}},\ }\href {\doibase 10.1103/PhysRevD.43.3191} {\bibfield  {journal}
  {\bibinfo  {journal} {Phys. Rev. D}\ }\textbf {\bibinfo {volume} {43}},\
  \bibinfo {pages} {3191} (\bibinfo {year} {1991})}\BibitemShut {NoStop}%
\bibitem [{\citenamefont {Aghanim}\ \emph {et~al.}(2020)\citenamefont {Aghanim}
  \emph {et~al.}}]{Aghanim:2018eyx}%
  \BibitemOpen
  \bibfield  {author} {\bibinfo {author} {\bibfnamefont {N.}~\bibnamefont
  {Aghanim}} \emph {et~al.} (\bibinfo {collaboration} {Planck}),\ }\href
  {\doibase 10.1051/0004-6361/201833910} {\bibfield  {journal} {\bibinfo
  {journal} {Astron. Astrophys.}\ }\textbf {\bibinfo {volume} {641}},\ \bibinfo
  {pages} {A6} (\bibinfo {year} {2020})},\ \Eprint
  {http://arxiv.org/abs/1807.06209} {arXiv:1807.06209 [astro-ph.CO]}
  \BibitemShut {NoStop}%
\bibitem [{\citenamefont {B\'elanger}\ \emph {et~al.}(2018)\citenamefont
  {B\'elanger}, \citenamefont {Boudjema}, \citenamefont {Goudelis},
  \citenamefont {Pukhov},\ and\ \citenamefont {Zaldivar}}]{Belanger:2018ccd}%
  \BibitemOpen
  \bibfield  {author} {\bibinfo {author} {\bibfnamefont {G.}~\bibnamefont
  {B\'elanger}}, \bibinfo {author} {\bibfnamefont {F.}~\bibnamefont
  {Boudjema}}, \bibinfo {author} {\bibfnamefont {A.}~\bibnamefont {Goudelis}},
  \bibinfo {author} {\bibfnamefont {A.}~\bibnamefont {Pukhov}}, \ and\ \bibinfo
  {author} {\bibfnamefont {B.}~\bibnamefont {Zaldivar}},\ }\href {\doibase
  10.1016/j.cpc.2018.04.027} {\bibfield  {journal} {\bibinfo  {journal}
  {Comput. Phys. Commun.}\ }\textbf {\bibinfo {volume} {231}},\ \bibinfo
  {pages} {173} (\bibinfo {year} {2018})},\ \Eprint
  {http://arxiv.org/abs/1801.03509} {arXiv:1801.03509 [hep-ph]} \BibitemShut
  {NoStop}%
\bibitem [{\citenamefont {Long}\ and\ \citenamefont {Van}(1999)}]{Long:1999ij}%
  \BibitemOpen
  \bibfield  {author} {\bibinfo {author} {\bibfnamefont {H.~N.}\ \bibnamefont
  {Long}}\ and\ \bibinfo {author} {\bibfnamefont {V.~T.}\ \bibnamefont {Van}},\
  }\href {\doibase 10.1088/0954-3899/25/12/302} {\bibfield  {journal} {\bibinfo
   {journal} {J. Phys. G}\ }\textbf {\bibinfo {volume} {25}},\ \bibinfo {pages}
  {2319} (\bibinfo {year} {1999})},\ \Eprint
  {http://arxiv.org/abs/hep-ph/9909302} {arXiv:hep-ph/9909302} \BibitemShut
  {NoStop}%
\bibitem [{\citenamefont {Belyaev}\ \emph {et~al.}(2013)\citenamefont
  {Belyaev}, \citenamefont {Christensen},\ and\ \citenamefont
  {Pukhov}}]{Belyaev:2012qa}%
  \BibitemOpen
  \bibfield  {author} {\bibinfo {author} {\bibfnamefont {A.}~\bibnamefont
  {Belyaev}}, \bibinfo {author} {\bibfnamefont {N.~D.}\ \bibnamefont
  {Christensen}}, \ and\ \bibinfo {author} {\bibfnamefont {A.}~\bibnamefont
  {Pukhov}},\ }\href {\doibase 10.1016/j.cpc.2013.01.014} {\bibfield  {journal}
  {\bibinfo  {journal} {Comput. Phys. Commun.}\ }\textbf {\bibinfo {volume}
  {184}},\ \bibinfo {pages} {1729} (\bibinfo {year} {2013})},\ \Eprint
  {http://arxiv.org/abs/1207.6082} {arXiv:1207.6082 [hep-ph]} \BibitemShut
  {NoStop}%
\bibitem [{\citenamefont {Liu}(1994)}]{Liu:1994rx}%
  \BibitemOpen
  \bibfield  {author} {\bibinfo {author} {\bibfnamefont {J.~T.}\ \bibnamefont
  {Liu}},\ }\href {\doibase 10.1103/PhysRevD.50.542} {\bibfield  {journal}
  {\bibinfo  {journal} {Phys. Rev. D}\ }\textbf {\bibinfo {volume} {50}},\
  \bibinfo {pages} {542} (\bibinfo {year} {1994})},\ \Eprint
  {http://arxiv.org/abs/hep-ph/9312312} {arXiv:hep-ph/9312312} \BibitemShut
  {NoStop}%
\bibitem [{\citenamefont {Rodriguez}\ and\ \citenamefont
  {Sher}(2004)}]{Rodriguez:2004mw}%
  \BibitemOpen
  \bibfield  {author} {\bibinfo {author} {\bibfnamefont {J.~A.}\ \bibnamefont
  {Rodriguez}}\ and\ \bibinfo {author} {\bibfnamefont {M.}~\bibnamefont
  {Sher}},\ }\href {\doibase 10.1103/PhysRevD.70.117702} {\bibfield  {journal}
  {\bibinfo  {journal} {Phys. Rev. D}\ }\textbf {\bibinfo {volume} {70}},\
  \bibinfo {pages} {117702} (\bibinfo {year} {2004})},\ \Eprint
  {http://arxiv.org/abs/hep-ph/0407248} {arXiv:hep-ph/0407248} \BibitemShut
  {NoStop}%
\bibitem [{\citenamefont {Benavides}\ \emph {et~al.}(2009)\citenamefont
  {Benavides}, \citenamefont {Giraldo},\ and\ \citenamefont
  {Ponce}}]{Benavides:2009cn}%
  \BibitemOpen
  \bibfield  {author} {\bibinfo {author} {\bibfnamefont {R.~H.}\ \bibnamefont
  {Benavides}}, \bibinfo {author} {\bibfnamefont {Y.}~\bibnamefont {Giraldo}},
  \ and\ \bibinfo {author} {\bibfnamefont {W.~A.}\ \bibnamefont {Ponce}},\
  }\href {\doibase 10.1103/PhysRevD.80.113009} {\bibfield  {journal} {\bibinfo
  {journal} {Phys. Rev. D}\ }\textbf {\bibinfo {volume} {80}},\ \bibinfo
  {pages} {113009} (\bibinfo {year} {2009})},\ \Eprint
  {http://arxiv.org/abs/0911.3568} {arXiv:0911.3568 [hep-ph]} \BibitemShut
  {NoStop}%
\bibitem [{\citenamefont {Cabarcas}\ \emph {et~al.}(2010)\citenamefont
  {Cabarcas}, \citenamefont {Gomez~Dumm},\ and\ \citenamefont
  {Martinez}}]{Cabarcas:2009vb}%
  \BibitemOpen
  \bibfield  {author} {\bibinfo {author} {\bibfnamefont {J.~M.}\ \bibnamefont
  {Cabarcas}}, \bibinfo {author} {\bibfnamefont {D.}~\bibnamefont
  {Gomez~Dumm}}, \ and\ \bibinfo {author} {\bibfnamefont {R.}~\bibnamefont
  {Martinez}},\ }\href {\doibase 10.1088/0954-3899/37/4/045001} {\bibfield
  {journal} {\bibinfo  {journal} {J. Phys. G}\ }\textbf {\bibinfo {volume}
  {37}},\ \bibinfo {pages} {045001} (\bibinfo {year} {2010})},\ \Eprint
  {http://arxiv.org/abs/0910.5700} {arXiv:0910.5700 [hep-ph]} \BibitemShut
  {NoStop}%
\bibitem [{\citenamefont {Cabarcas}\ \emph {et~al.}(2012)\citenamefont
  {Cabarcas}, \citenamefont {Duarte},\ and\ \citenamefont
  {Rodriguez}}]{Cabarcas:2011hb}%
  \BibitemOpen
  \bibfield  {author} {\bibinfo {author} {\bibfnamefont {J.~M.}\ \bibnamefont
  {Cabarcas}}, \bibinfo {author} {\bibfnamefont {J.}~\bibnamefont {Duarte}}, \
  and\ \bibinfo {author} {\bibfnamefont {J.-A.}\ \bibnamefont {Rodriguez}},\
  }\href {\doibase 10.1155/2012/657582} {\bibfield  {journal} {\bibinfo
  {journal} {Adv. High Energy Phys.}\ }\textbf {\bibinfo {volume} {2012}},\
  \bibinfo {pages} {657582} (\bibinfo {year} {2012})},\ \Eprint
  {http://arxiv.org/abs/1111.0315} {arXiv:1111.0315 [hep-ph]} \BibitemShut
  {NoStop}%
\bibitem [{\citenamefont {Cogollo}\ \emph {et~al.}(2012)\citenamefont
  {Cogollo}, \citenamefont {de~Andrade}, \citenamefont {Queiroz},\ and\
  \citenamefont {Rebello~Teles}}]{Cogollo:2012ek}%
  \BibitemOpen
  \bibfield  {author} {\bibinfo {author} {\bibfnamefont {D.}~\bibnamefont
  {Cogollo}}, \bibinfo {author} {\bibfnamefont {A.~V.}\ \bibnamefont
  {de~Andrade}}, \bibinfo {author} {\bibfnamefont {F.~S.}\ \bibnamefont
  {Queiroz}}, \ and\ \bibinfo {author} {\bibfnamefont {P.}~\bibnamefont
  {Rebello~Teles}},\ }\href {\doibase 10.1140/epjc/s10052-012-2029-7}
  {\bibfield  {journal} {\bibinfo  {journal} {Eur. Phys. J. C}\ }\textbf
  {\bibinfo {volume} {72}},\ \bibinfo {pages} {2029} (\bibinfo {year}
  {2012})},\ \Eprint {http://arxiv.org/abs/1201.1268} {arXiv:1201.1268
  [hep-ph]} \BibitemShut {NoStop}%
\bibitem [{\citenamefont {Machado}\ \emph {et~al.}(2013)\citenamefont
  {Machado}, \citenamefont {Montero},\ and\ \citenamefont
  {Pleitez}}]{Machado:2013jca}%
  \BibitemOpen
  \bibfield  {author} {\bibinfo {author} {\bibfnamefont {A.~C.~B.}\
  \bibnamefont {Machado}}, \bibinfo {author} {\bibfnamefont {J.~C.}\
  \bibnamefont {Montero}}, \ and\ \bibinfo {author} {\bibfnamefont
  {V.}~\bibnamefont {Pleitez}},\ }\href {\doibase 10.1103/PhysRevD.88.113002}
  {\bibfield  {journal} {\bibinfo  {journal} {Phys. Rev. D}\ }\textbf {\bibinfo
  {volume} {88}},\ \bibinfo {pages} {113002} (\bibinfo {year} {2013})},\
  \Eprint {http://arxiv.org/abs/1305.1921} {arXiv:1305.1921 [hep-ph]}
  \BibitemShut {NoStop}%
\bibitem [{\citenamefont {Taoso}\ \emph {et~al.}(2008)\citenamefont {Taoso},
  \citenamefont {Bertone},\ and\ \citenamefont {Masiero}}]{Taoso_2008}%
  \BibitemOpen
  \bibfield  {author} {\bibinfo {author} {\bibfnamefont {M.}~\bibnamefont
  {Taoso}}, \bibinfo {author} {\bibfnamefont {G.}~\bibnamefont {Bertone}}, \
  and\ \bibinfo {author} {\bibfnamefont {A.}~\bibnamefont {Masiero}},\ }\href
  {\doibase 10.1088/1475-7516/2008/03/022} {\bibfield  {journal} {\bibinfo
  {journal} {Journal of Cosmology and Astroparticle Physics}\ }\textbf
  {\bibinfo {volume} {2008}},\ \bibinfo {pages} {022} (\bibinfo {year}
  {2008})}\BibitemShut {NoStop}%
\bibitem [{\citenamefont {Hooper}(2010)}]{Hooper:2009zm}%
  \BibitemOpen
  \bibfield  {author} {\bibinfo {author} {\bibfnamefont {D.}~\bibnamefont
  {Hooper}},\ }in\ \href {\doibase 10.1142/9789812838360_0014} {\emph {\bibinfo
  {booktitle} {{Theoretical Advanced Study Institute in Elementary Particle
  Physics}: {The Dawn of the LHC Era}}}}\ (\bibinfo {year} {2010})\ pp.\
  \bibinfo {pages} {709--764},\ \Eprint {http://arxiv.org/abs/0901.4090}
  {arXiv:0901.4090 [hep-ph]} \BibitemShut {NoStop}%
\bibitem [{\citenamefont {Munoz}(2004)}]{Munoz:2003gx}%
  \BibitemOpen
  \bibfield  {author} {\bibinfo {author} {\bibfnamefont {C.}~\bibnamefont
  {Munoz}},\ }\href {\doibase 10.1142/S0217751X04018154} {\bibfield  {journal}
  {\bibinfo  {journal} {Int. J. Mod. Phys. A}\ }\textbf {\bibinfo {volume}
  {19}},\ \bibinfo {pages} {3093} (\bibinfo {year} {2004})},\ \Eprint
  {http://arxiv.org/abs/hep-ph/0309346} {arXiv:hep-ph/0309346} \BibitemShut
  {NoStop}%
\bibitem [{\citenamefont {Bertone}\ \emph {et~al.}(2005)\citenamefont
  {Bertone}, \citenamefont {Hooper},\ and\ \citenamefont
  {Silk}}]{BERTONE2005279}%
  \BibitemOpen
  \bibfield  {author} {\bibinfo {author} {\bibfnamefont {G.}~\bibnamefont
  {Bertone}}, \bibinfo {author} {\bibfnamefont {D.}~\bibnamefont {Hooper}}, \
  and\ \bibinfo {author} {\bibfnamefont {J.}~\bibnamefont {Silk}},\ }\href
  {\doibase https://doi.org/10.1016/j.physrep.2004.08.031} {\bibfield
  {journal} {\bibinfo  {journal} {Physics Reports}\ }\textbf {\bibinfo {volume}
  {405}},\ \bibinfo {pages} {279} (\bibinfo {year} {2005})}\BibitemShut
  {NoStop}%
\bibitem [{\citenamefont {Gascon}(2004)}]{GASCON200496}%
  \BibitemOpen
  \bibfield  {author} {\bibinfo {author} {\bibfnamefont {J.}~\bibnamefont
  {Gascon}},\ }\href {\doibase https://doi.org/10.1016/j.nima.2003.11.251}
  {\bibfield  {journal} {\bibinfo  {journal} {Nuclear Instruments and Methods
  in Physics Research Section A: Accelerators, Spectrometers, Detectors and
  Associated Equipment}\ }\textbf {\bibinfo {volume} {520}},\ \bibinfo {pages}
  {96} (\bibinfo {year} {2004})},\ \bibinfo {note} {proceedings of the 10th
  International Workshop on Low Temperature Detectors}\BibitemShut {NoStop}%
\bibitem [{\citenamefont {Ramachers}(2003)}]{RAMACHERS2003341}%
  \BibitemOpen
  \bibfield  {author} {\bibinfo {author} {\bibfnamefont {Y.}~\bibnamefont
  {Ramachers}},\ }\href {\doibase
  https://doi.org/10.1016/S0920-5632(03)01327-6} {\bibfield  {journal}
  {\bibinfo  {journal} {Nuclear Physics B - Proceedings Supplements}\ }\textbf
  {\bibinfo {volume} {118}},\ \bibinfo {pages} {341} (\bibinfo {year}
  {2003})},\ \bibinfo {note} {proceedings of the XXth International Conference
  on Neutrino Physics and Astrophysics}\BibitemShut {NoStop}%
\bibitem [{\citenamefont {Shan}(2007)}]{Shan:2007mq}%
  \BibitemOpen
  \bibfield  {author} {\bibinfo {author} {\bibfnamefont {C.-L.}\ \bibnamefont
  {Shan}},\ }\emph {\bibinfo {title} {{Theoretical Interpretation of
  Experimental Data from Direct Dark Matter Detection}}},\ \href@noop {}
  {\bibinfo {type} {Other thesis}} (\bibinfo {year} {2007}),\ \Eprint
  {http://arxiv.org/abs/0707.0488} {arXiv:0707.0488 [astro-ph]} \BibitemShut
  {NoStop}%
\bibitem [{\citenamefont {Ramirez~Barreto}\ \emph {et~al.}(2010)\citenamefont
  {Ramirez~Barreto}, \citenamefont {Coutinho},\ and\ \citenamefont
  {Sa~Borges}}]{RamirezBarreto:2010vji}%
  \BibitemOpen
  \bibfield  {author} {\bibinfo {author} {\bibfnamefont {E.}~\bibnamefont
  {Ramirez~Barreto}}, \bibinfo {author} {\bibfnamefont {Y.~A.}\ \bibnamefont
  {Coutinho}}, \ and\ \bibinfo {author} {\bibfnamefont {J.}~\bibnamefont
  {Sa~Borges}},\ }\href {\doibase 10.1016/j.physletb.2010.04.039} {\bibfield
  {journal} {\bibinfo  {journal} {Phys. Lett. B}\ }\textbf {\bibinfo {volume}
  {689}},\ \bibinfo {pages} {36} (\bibinfo {year} {2010})},\ \Eprint
  {http://arxiv.org/abs/1004.3269} {arXiv:1004.3269 [hep-ph]} \BibitemShut
  {NoStop}%
\bibitem [{\citenamefont {Ramirez~Barreto}\ \emph {et~al.}(2007)\citenamefont
  {Ramirez~Barreto}, \citenamefont {do~Amaral~Coutinho},\ and\ \citenamefont
  {Sa~Borges}}]{RamirezBarreto:2007cie}%
  \BibitemOpen
  \bibfield  {author} {\bibinfo {author} {\bibfnamefont {E.}~\bibnamefont
  {Ramirez~Barreto}}, \bibinfo {author} {\bibfnamefont {Y.}~\bibnamefont
  {do~Amaral~Coutinho}}, \ and\ \bibinfo {author} {\bibfnamefont
  {J.}~\bibnamefont {Sa~Borges}},\ }\href {\doibase
  10.1140/epjc/s10052-007-0254-2} {\bibfield  {journal} {\bibinfo  {journal}
  {Eur. Phys. J. C}\ }\textbf {\bibinfo {volume} {50}},\ \bibinfo {pages} {909}
  (\bibinfo {year} {2007})},\ \Eprint {http://arxiv.org/abs/hep-ph/0703099}
  {arXiv:hep-ph/0703099} \BibitemShut {NoStop}%
\bibitem [{\citenamefont {Ramirez~Barreto}\ \emph {et~al.}(2006)\citenamefont
  {Ramirez~Barreto}, \citenamefont {do~Amaral~Coutinho},\ and\ \citenamefont
  {Sa~Borges}}]{RamirezBarreto:2006tn}%
  \BibitemOpen
  \bibfield  {author} {\bibinfo {author} {\bibfnamefont {E.}~\bibnamefont
  {Ramirez~Barreto}}, \bibinfo {author} {\bibfnamefont {Y.}~\bibnamefont
  {do~Amaral~Coutinho}}, \ and\ \bibinfo {author} {\bibfnamefont
  {J.}~\bibnamefont {Sa~Borges}},\ }\href@noop {} {\  (\bibinfo {year}
  {2006})},\ \Eprint {http://arxiv.org/abs/hep-ph/0605098}
  {arXiv:hep-ph/0605098} \BibitemShut {NoStop}%
\bibitem [{\citenamefont {Coutinho}\ \emph {et~al.}(2013)\citenamefont
  {Coutinho}, \citenamefont {Salustino Guimar\~aes},\ and\ \citenamefont
  {Nepomuceno}}]{Coutinho:2013lta}%
  \BibitemOpen
  \bibfield  {author} {\bibinfo {author} {\bibfnamefont {Y.~A.}\ \bibnamefont
  {Coutinho}}, \bibinfo {author} {\bibfnamefont {V.}~\bibnamefont {Salustino
  Guimar\~aes}}, \ and\ \bibinfo {author} {\bibfnamefont {A.~A.}\ \bibnamefont
  {Nepomuceno}},\ }\href {\doibase 10.1103/PhysRevD.87.115014} {\bibfield
  {journal} {\bibinfo  {journal} {Phys. Rev. D}\ }\textbf {\bibinfo {volume}
  {87}},\ \bibinfo {pages} {115014} (\bibinfo {year} {2013})},\ \Eprint
  {http://arxiv.org/abs/1304.7907} {arXiv:1304.7907 [hep-ph]} \BibitemShut
  {NoStop}%
\bibitem [{\citenamefont {Long}\ and\ \citenamefont
  {Inami}(2000)}]{Long:1999bny}%
  \BibitemOpen
  \bibfield  {author} {\bibinfo {author} {\bibfnamefont {H.~N.}\ \bibnamefont
  {Long}}\ and\ \bibinfo {author} {\bibfnamefont {T.}~\bibnamefont {Inami}},\
  }\href {\doibase 10.1103/PhysRevD.61.075002} {\bibfield  {journal} {\bibinfo
  {journal} {Phys. Rev. D}\ }\textbf {\bibinfo {volume} {61}},\ \bibinfo
  {pages} {075002} (\bibinfo {year} {2000})},\ \Eprint
  {http://arxiv.org/abs/hep-ph/9902475} {arXiv:hep-ph/9902475} \BibitemShut
  {NoStop}%
\bibitem [{\citenamefont {Alves}\ \emph {et~al.}(2022)\citenamefont {Alves},
  \citenamefont {Duarte}, \citenamefont {Kovalenko}, \citenamefont
  {Oviedo-Torres}, \citenamefont {Queiroz},\ and\ \citenamefont
  {Villamizar}}]{Alves:2022hcp}%
  \BibitemOpen
  \bibfield  {author} {\bibinfo {author} {\bibfnamefont {A.}~\bibnamefont
  {Alves}}, \bibinfo {author} {\bibfnamefont {L.}~\bibnamefont {Duarte}},
  \bibinfo {author} {\bibfnamefont {S.}~\bibnamefont {Kovalenko}}, \bibinfo
  {author} {\bibfnamefont {Y.~M.}\ \bibnamefont {Oviedo-Torres}}, \bibinfo
  {author} {\bibfnamefont {F.~S.}\ \bibnamefont {Queiroz}}, \ and\ \bibinfo
  {author} {\bibfnamefont {Y.~S.}\ \bibnamefont {Villamizar}},\ }\href@noop {}
  {\  (\bibinfo {year} {2022})},\ \Eprint {http://arxiv.org/abs/2203.02520}
  {arXiv:2203.02520 [hep-ph]} \BibitemShut {NoStop}%
\bibitem [{\citenamefont {Griest}\ and\ \citenamefont
  {Kamionkowski}(1990)}]{Griest:1989wd}%
  \BibitemOpen
  \bibfield  {author} {\bibinfo {author} {\bibfnamefont {K.}~\bibnamefont
  {Griest}}\ and\ \bibinfo {author} {\bibfnamefont {M.}~\bibnamefont
  {Kamionkowski}},\ }\href {\doibase 10.1103/PhysRevLett.64.615} {\bibfield
  {journal} {\bibinfo  {journal} {Phys. Rev. Lett.}\ }\textbf {\bibinfo
  {volume} {64}},\ \bibinfo {pages} {615} (\bibinfo {year} {1990})}\BibitemShut
  {NoStop}%
\end{thebibliography}%

\end{document}